\documentstyle[12pt,epsf]{article}

\newlength{\pubnumber} 

\catcode`\@=11
\@addtoreset{equation}{section}

\def\section{\@startsection{section}{1}{\z@}{3.5ex plus 1ex minus .2ex}
 {2.3ex plus .2ex}{\large\bf}}
\def\subsection{\@startsection{subsection}{2}{\z@}{2.3ex plus .2ex}
 {2.3ex plus .2ex}{\bf}}

\begin{document}

\begin{titlepage}
\samepage{
\setcounter{page}{1}
\rightline{ACT-1/00}
\rightline{CTP-TAMU-02/00}
\rightline{OUTP--00--03P}
\rightline{TPI-MINN-00/06}
\rightline{UMN--TH--1842-00}
\rightline{\tt hep-ph/0002060}
\rightline{January 2000}
\vfill
\begin{center}
 {\Large \bf Non--Abelian Flat Directions in a\\ 
   Minimal Superstring Standard Model}
\vfill
\vskip .4truecm
\vfill {\large
        G.B. Cleaver,$^{1,2}$\footnote{gcleaver@rainbow.physics.tamu.edu}
        A.E. Faraggi,$^{3}$\footnote{faraggi@mnhepw.hep.umn.edu}
        D.V. Nanopoulos$^{1,2,4}$\footnote{dimitri@soda.physics.tamu.edu},}

       {\large and
        J.W. Walker$^{1}$\footnote{jwalker@rainbow.physics.tamu.edu}}
\\
\vspace{.12in}
{\it $^{1}$ Center for Theoretical Physics,
            Dept.\  of Physics, Texas A\&M University,\\
            College Station, TX 77843, USA\\}
\vspace{.06in}
{\it $^{2}$ Astro Particle Physics Group,
            Houston Advanced Research Center (HARC),\\
            The Mitchell Campus,
            Woodlands, TX 77381, USA\\}
\vspace{.06in}
{\it$^{3}$  Department of Physics, University of Minnesota, 
            Minneapolis, MN 55455, USA,\\
	and Theoretical Physics, University of Oxford, 1 Keble Road,
Oxford OX1 3NP, UK\\}
\vspace{.025in}
{\it$^{4}$  Academy of Athens, Chair of Theoretical Physics, 
            Division of Natural Sciences,\\
            28 Panepistimiou Avenue, Athens 10679, Greece\\}
\vspace{.025in}
\end{center}
\vfill
\begin{abstract}
Recently, by studying exact flat directions of non--Abelian singlet fields,
we demonstrated the existence of free fermionic heterotic--string models
in which the $SU(3)_C\times SU(2)_L\times U(1)_Y$--charged matter spectrum,
just below the string scale, consists solely of the MSSM spectrum. 
In this paper we generalize the analysis to include VEVs of non--Abelian
fields. We find several, MSSM--producing, exact non--Abelian flat directions,
which are the first such examples in the literature. We 
examine the possibility that hidden sector condensates lift the flat
directions. 
\end{abstract}
\smallskip}
\end{titlepage}

\setcounter{footnote}{0}

\def\at{ }
\def\beq{\begin{equation}}
\def\eeq{\end{equation}}
\def\beqn{\begin{eqnarray}}
\def\eeqn{\end{eqnarray}}
\def\no{\noindent }
\def\nolabel{\nonumber }

\def\NA{non--Abelian }

\def\gsim{{\buildrel >\over \sim}}
\def\lsim{{\buildrel <\over \sim}}

\def\ie{i.e., }
\def\eg{{\it e.g.}}
\def\eq#1{eq.\ (\ref{#1})}

\def\lt{<}

\def\slash#1{#1\hskip-6pt/\hskip6pt}
\def\slk{\slash{k}}

\def\dag{\dagger}
\def\qandq{\quad {\rm and} \quad} 
\def\qand{\quad {\rm and} } 
\def\andq{ {\rm and} \quad } 
\def\qwithq{\quad {\rm with} \quad} 
\def\qwith{ \quad {\rm with} } 
\def\withq{ {\rm with} \quad} 

\def\fhalf{\frac{1}{2}}
\def\fsqrt{\frac{1}{\sqrt{2}}}
\def\half{{\textstyle{1\over 2}}}
\def\third{{\textstyle {1\over3}}}
\def\quarter{{\textstyle {1\over4}}}
\def\sixth{{\textstyle {1\over6}}}
\def\m{$\phantom{-}$}
\def\j{$-$}
\def\ps{{\tt +}}
\def\pps{\phantom{+}}

\def\zz{$Z_2\times Z_2$ }

\def\Tr{{\rm Tr}\, }
\def\tr{{\rm tr}\, }

\def\MP{M_{P}}
\def\GeV{\,{\rm GeV}}
\def\TeV{\,{\rm TeV}}

\def\lam#1{\lambda_{#1}}
\def\non{\nonumber}
\def\smgg{ $SU(3)_C\times SU(2)_L\times U(1)_Y$ }
\def\smggb{ $SU(3)_C\times SU(2)_L\times U(1)_Y$}
\def\SM{Standard--Model }
\def\SUSY{supersymmetry }
\def\SSSM{supersymmetric standard model}
\def\MSSM{minimal supersymmetric standard model}
\def\MSSSM{MS$_{str}$SM }
\def\MSSSMc{MS$_{str}$SM, }
\def\obs{{\rm observable}}
\def\sig{{\rm singlets}}
\def\hid{{\rm hidden}}
\def\MS{M_{str}}
\def\Ms{$M_{str}$}
\def\MP{M_{P}}

\def\vev#1{\langle #1\rangle}
\def\mvev#1{|\langle #1\rangle|^2}

\def\UA{U(1)_{\rm A}}
\def\QA{Q^{(\rm A)}}
\def\mssm{SU(3)_C\times SU(2)_L\times U(1)_Y} 

\def\KM{Ka\v c--Moody }

\def\y{\,{\rm y}}
\def\l{\langle}
\def\r{\rangle}
\def\o#1{\frac{1}{#1}}

\def\zi{z_{\infty}}

\def\hb#1{\bar{h}_{#1}}
\def\Htw{{\tilde H}}
\def\chibar{{\overline{\chi}}}
\def\qbar{{\overline{q}}}
\def\ibar{{\overline{\imath}}}
\def\jbar{{\overline{\jmath}}}
\def\Hbar{{\overline{H}}}
\def\Qbar{{\overline{Q}}}
\def\abar{{\overline{a}}}
\def\alphabar{{\overline{\alpha}}}
\def\betabar{{\overline{\beta}}}
\def\tautwo{{ \tau_2 }}
\def\thetatwo{{ \vartheta_2 }}
\def\thetathree{{ \vartheta_3 }}
\def\thetafour{{ \vartheta_4 }}
\def\ttwo{{\vartheta_2}}
\def\tthree{{\vartheta_3}}
\def\tfour{{\vartheta_4}}
\def\ti{{\vartheta_i}}
\def\tj{{\vartheta_j}}
\def\tk{{\vartheta_k}}
\def\calF{{\cal F}}
\def\smallmatrix#1#2#3#4{{ {{#1}~{#2}\choose{#3}~{#4}} }}
\def\ab{{\alpha\beta}}
\def\Minv{{ (M^{-1}_\ab)_{ij} }}
\def\ii{{(i)}}
\def\V{{\bf V}}
\def\N{{\bf N}}

\def\b{{\bf b}}
\def\S{{\bf S}}
\def\X{{\bf X}}
\def\I{{\bf I}}
\def\bone{{\mathbf 1}}
\def\bo{{\mathbf 0}}
\def\bs{{\mathbf S}}
\def\mS{{\mathbf S}}
\def\bS{{\mathbf S}}
\def\bb{{\mathbf b}}
\def\mb{{\mathbf b}}
\def\mX{{\mathbf X}}
\def\mI{{\mathbf I}}
\def\bI{{\mathbf I}}
\def\balpha{{\mathbf \alpha}}
\def\bbeta{{\mathbf \beta}}
\def\bgamma{{\mathbf \gamma}}
\def\bxi{{\mathbf \xi}}
\def\malpha{{\mathbf \alpha}}
\def\mbeta{{\mathbf \beta}}
\def\mgamma{{\mathbf \gamma}}
\def\mxi{{\mathbf \xi}}
\def\bphi{\overline{\Phi}}

\def\eps{\epsilon}

\def\t#1#2{{ \Theta\left\lbrack \matrix{ {#1}\cr {#2}\cr }\right\rbrack }}
\def\C#1#2{{ C\left\lbrack \matrix{ {#1}\cr {#2}\cr }\right\rbrack }}
\def\tp#1#2{{ \Theta'\left\lbrack \matrix{ {#1}\cr {#2}\cr }\right\rbrack }}
\def\tpp#1#2{{ \Theta''\left\lbrack \matrix{ {#1}\cr {#2}\cr }\right\rbrack }}
\def\l{\langle}
\def\r{\rangle}

\def\op#1{$\Phi_{#1}$}
\def\opp#1{$\Phi^{'}_{#1}$}
\def\opb#1{$\overline{\Phi}_{#1}$}
\def\opbp#1{$\overline{\Phi}^{'}_{#1}$}
\def\oppb#1{$\overline{\Phi}^{'}_{#1}$}
\def\oppx#1{$\Phi^{(')}_{#1}$}
\def\opbpx#1{$\overline{\Phi}^{(')}_{#1}$}

\def\oh#1{$h_{#1}$}
\def\ohb#1{${\bar{h}}_{#1}$}
\def\ohp#1{$h^{'}_{#1}$}

\def\oQ#1{$Q_{#1}$}
\def\odc#1{$d^{c}_{#1}$}
\def\ouc#1{$u^{c}_{#1}$}

\def\oL#1{$L_{#1}$}
\def\oec#1{$e^{c}_{#1}$}
\def\oNc#1{$N^{c}_{#1}$}

\def\oH#1{$H_{#1}$}
\def\oV#1{$V_{#1}$}
\def\oHs#1{$H^{s}_{#1}$}
\def\oVs#1{$V^{s}_{#1}$}

\def\p#1{{\Phi_{#1}}}
\def\pp#1{{\Phi^{'}_{#1}}}
\def\pb#1{{{\overline{\Phi}}_{#1}}}
\def\pbp#1{{{\overline{\Phi}}^{'}_{#1}}}
\def\ppb#1{{{\overline{\Phi}}^{'}_{#1}}}
\def\ppx#1{{\Phi^{(')}_{#1}}}
\def\pbpx#1{{\overline{\Phi}^{(')}_{#1}}}

\def\h#1{h_{#1}}
\def\hb#1{{\bar{h}}_{#1}}
\def\hp#1{h^{'}_{#1}}

\def\Q#1{Q_{#1}}
\def\dc#1{d^{c}_{#1}}
\def\uc#1{u^{c}_{#1}}

\def\L#1{L_{#1}}
\def\ec#1{e^{c}_{#1}}
\def\Nc#1{N^{c}_{#1}}

\def\H#1{H_{#1}}
\def\V#1{V_{#1}}
\def\Hs#1{{H^{s}_{#1}}}
\def\Vs#1{{V^{s}_{#1}}}

\def\fdtv{FD2V }
\def\fdtp{FD2$^{'}$ }
\def\fdtpv{FD2$^{'}$v }

\def\FD2pv{FD2$^{'}$V }
\def\FD2p{FD2$^{'}$ }


\def\inbar{\,\vrule height1.5ex width.4pt depth0pt}

\def\IC{\relax\hbox{$\inbar\kern-.3em{\rm C}$}}
\def\IQ{\relax\hbox{$\inbar\kern-.3em{\rm Q}$}}
\def\IR{\relax{\rm I\kern-.18em R}}
 \font\cmss=cmss10 \font\cmsss=cmss10 at 7pt
 \font\cmsst=cmss10 at 9pt
 \font\cmssn=cmss9

\def\IZ{\relax\ifmmode\mathchoice
 {\hbox{\cmss Z\kern-.4em Z}}{\hbox{\cmss Z\kern-.4em Z}}
 {\lower.9pt\hbox{\cmsss Z\kern-.4em Z}}
 {\lower1.2pt\hbox{\cmsss Z\kern-.4em Z}}\else{\cmss Z\kern-.4em Z}\fi}

\def\AEF{A.E. Faraggi}
\def\AP#1#2#3{{\it Ann.\ Phys.}\/ {\bf#1} (#2) #3}
\def\NPB#1#2#3{{\it Nucl.\ Phys.}\/ {\bf B#1} (#2) #3}
\def\NPBPS#1#2#3{{\it Nucl.\ Phys.}\/ {{\bf B} (Proc. Suppl.) {\bf #1}} (#2) 
 #3}
\def\PLB#1#2#3{{\it Phys.\ Lett.}\/ {\bf B#1} (#2) #3}
\def\PRD#1#2#3{{\it Phys.\ Rev.}\/ {\bf D#1} (#2) #3}
\def\PRL#1#2#3{{\it Phys.\ Rev.\ Lett.}\/ {\bf #1} (#2) #3}
\def\PRT#1#2#3{{\it Phys.\ Rep.}\/ {\bf#1} (#2) #3}
\def\PTP#1#2#3{{\it Prog.\ Theo.\ Phys.}\/ {\bf#1} (#2) #3}
\def\MODA#1#2#3{{\it Mod.\ Phys.\ Lett.}\/ {\bf A#1} (#2) #3}
\def\IJMP#1#2#3{{\it Int.\ J.\ Mod.\ Phys.}\/ {\bf A#1} (#2) #3}
\def\nuvc#1#2#3{{\it Nuovo Cimento}\/ {\bf #1A} (#2) #3}
\def\RPP#1#2#3{{\it Rept.\ Prog.\ Phys.}\/ {\bf #1} (#2) #3}
\def\etal{{\it et al\/}}

\hyphenation{su-per-sym-met-ric non-su-per-sym-met-ric}
\hyphenation{space-time-super-sym-met-ric}
\hyphenation{mod-u-lar mod-u-lar--in-var-i-ant}


\section{Minimal Superstring Standard Models} 

The most realistic string models found to date \cite{fny,real} have been constructed
in the free fermionic formulation \cite{fff} of the heterotic--string.
A large number of three generation models, which differ in their
detailed phenomenological characteristics, have been built.
All these models share an underlying $Z_2\times Z_2$ orbifold 
structure, which naturally gives rise to three generations
with the $SO(10)$ embedding \cite{nahe} of the Standard Model 
spectrum\footnote{Among the three generation orbifold models,
constructed to date, only the free fermionic models possess
the $SO(10)$ embedding of the Standard Model spectrum}.
Recently, 
it was further demonstrated that free fermionic heterotic--string 
models can also produce models with solely the 
spectrum of the Minimal Supersymmetric Standard Model (MSSM)
in the effective low energy field theory \cite{cfn1,cfn2,cfn3}. 
This is achieved due to the decoupling of all non--MSSM,
exotic, and not--exotic, string states, at or slightly
below the string scale, by Standard Model singlet VEVs
which cancel the anomalous $U(1)$ D--term. This provides,
for the first time, an example of a Minimal Standard
Heterotic--String Model (MSHSM).

The emergence of a MSHSM in the free fermionic formulation
reinforces the motivation for an improved understanding
of this class of string compactifications.
One of the important advancements of the last few years
has been the development of techniques for 
systematic analysis of the $F$-- and $D$--flat directions 
of (string) models. 
Indeed, in demonstrating
the existence of a free fermionic MSHSM we have utilized those improved
techniques \cite{cfn1,cfn2,cfn3}. 
However, one limitation of those systematic 
studies performed to date is that they have
included only flat directions of non--Abelian singlet fields. 
That is, fields which are singlets of all the non--Abelian
gauge groups of a given string model and which may carry only
Abelian $U(1)$ charges, or are singlets of the entire 
four dimensional gauge group. On the other hand, 
it has been shown in the past that some of the phenomenological
constraints, such as quark--mixing \cite{namix}, may necessitate the 
use of non--Abelian VEVs. This was also suggested in our recent 
exploration of possible generational mass hierarchies and
effective Higgs $\mu$ terms resulting from singlet VEVs in our
MSHSM \cite{cfn3}.  

 In this letter we therefore 
begin the task of extending the systematic analysis 
of flat directions for the cases which include non--Abelian VEVs. 
For our investigation we again start with the model we have 
denoted ``FNY,'' first introduced in \cite{fny}. 
An important question that has been of some debate in previous
studies, and is relevant for the question of supersymmetry
breaking, is whether flat directions which include 
non--Abelian VEVs can be exact. Indeed, 
a particularly important result which we show here
for the first time is the demonstration of a MSHSM solution which
includes non--Abelian VEVs and is flat to all orders of
nonrenormalizable terms. We further elaborate on the 
specific complications which arise in considering non--Abelian
VEVs in the string models and briefly discuss some of the 
phenomenological implications of non--Abelian VEVs in the
MSHSM. In a follow up to this letter \cite{cfn5} we will present
a large collection of (systematically generated) 
non--Abelian MSHSM $D$--flat directions 
that retain $F$--flatness to at least seventh order, 
along with a study of the phenomenological features of these directions.

\section{Non--Abelian Flat MSSM directions of the FNY model}

As advertised above, the model that we choose to study in
this paper is the FNY model \cite{fny}, which produced
the first example of a MSHSM. The boundary conditions 
and GSO projection coefficients which define the model
are given in ref. \cite{fny,cfn1,cfn2,cfn3} together with the cubic 
level and higher order terms in the superpotential. 
Here, we plunge directly to the analysis of the 
non--Abelian flat directions.

\subsection{Generic $D$-- and $F$--Flatness Constraints}

Spacetime supersymmetry is broken in a model
when the expectation value of the scalar potential,
\beqn
 V(\varphi) = \half \sum_{\alpha} g_{\alpha} D_a^{\alpha} D_a^{\alpha} +
                    \sum_i | F_{\varphi_i} |^2\,\, ,
\label{vdef}
\eeqn
becomes non--zero. 
The $D$--term contributions in (\ref{vdef}) have the form,   
\beqn
D_a^{\alpha}&\equiv& \sum_m \varphi_{m}^{\dagger} T^{\alpha}_a \varphi_m\,\, , 
\label{dtgen} 
\eeqn
with $T^{\alpha}_a$ a matrix generator of the gauge group $g_{\alpha}$ 
for the representation $\varphi_m$, 
while the $F$--term contributions are, 
\beqn
F_{\Phi_{m}} &\equiv& \frac{\partial W}{\partial \Phi_{m}} \label{ftgen}\,\, . 
\eeqn
The $\varphi_m$ are the scalar field superpartners     
of the chiral spin--$\half$ fermions $\psi_m$, which together  
form a superfield $\Phi_{m}$.
Since all of the $D$ and $F$ contributions to (\ref{vdef}) 
are positive semidefinite, each must have 
a zero expectation value for supersymmetry to remain unbroken.

For an Abelian gauge group, the $D$--term (\ref{dtgen}) simplifies to
\beqn
D^{i}&\equiv& \sum_m  Q^{(i)}_m | \varphi_m |^2 \label{dtab}\,\,  
\eeqn
where $Q^{(i)}_m$ is the $U(1)_i$ charge of $\varphi_m$.  
When an Abelian symmetry is anomalous, that is,
the trace of its charge 
over the massless fields is non--zero, 
\beqn
\Tr Q^{(A)}\ne 0\,\, ,
\label{qtnz}
\eeqn 
the associated $D$--term acquires a Fayet--Iliopoulos (FI) term,
$\eps\equiv\frac{g^2_s M_P^2}{192\pi^2}\Tr Q^{(A)}$, 
\beqn
D^{(A)}&\equiv& \sum_m  Q^{(A)}_m | \varphi_m |^2 
+ \eps \, .
\label{dtaban}  
\eeqn  
$g_{s}$ is the string coupling and $M_P$ is the reduced Planck mass, 
$M_P\equiv M_{Planck}/\sqrt{8 \pi}\approx 2.4\times 10^{18}$ GeV. 

The FI term breaks supersymmetry near the string scale,
\beqn 
V \sim g_{s}^{2} \eps^2\,\, ,\label{veps}
\eeqn  
unless its can be cancelled by a set of scalar VEVs, $\{\vev{\varphi_{m'}}\}$, 
carrying anomalous charges $Q^{(A)}_{m'}$,
\beq
\vev{D^{(A)}}= \sum_{m'} Q^{(A)}_{m'} |\vev{\varphi_{m'}}|^2 
+ \eps  = 0\,\, .
\label{daf}
\eeq
To maintain supersymmetry, a set of anomaly--cancelling VEVs must 
simultaneously be $D$--flat 
for all additional Abelian and the non--Abelian gauge groups, 
\beq
\vev{D^{i,\alpha}}= 0\,\, . 
\label{dana}
\eeq

A non--trivial superpotential $W$ also imposes numerous constraints on allowed
sets of anomaly--cancelling VEVs, through the $F$--terms in (\ref{vdef}).
$F$--flatness (and thereby supersymmetry) can be broken through an 
$n^{\rm th}$--order $W$ term containing $\Phi_{m}$ when all of the additional 
fields in the term acquire VEVs,
\beqn
\vev{F_{\Phi_m}}&\sim& \vev{{\frac{\partial W}{\partial \Phi_{m}}}} 
         \sim \lambda_n \vev{\varphi}^2 (\frac{\vev{\varphi}}{\MS})^{n-3}\,\, ,
\label{fwnb2}
\eeqn
where $\varphi$ denotes a generic scalar VEV.
If $\Phi_{m}$ additionally has a VEV, then
supersymmetry can be broken simply by $\vev{W} \ne 0$.
(The lower the order of an $F$--breaking term, 
the closer the supersymmetry breaking scale 
is to the string scale.) 

\subsection{Non--Abelian flat directions in the FNY Model}

In \cite{cfn2} we classified the MSSM producing flat directions of the 
FNY model
that are composed solely of singlet fields. Following this, in \cite{cfn3}
we studied the phenomenological features of these singlet directions.
We now consider here generalized MSSM--producing
flat directions in the FNY model that contain \NA VEVs.
In our prior investigations we demanded stringent flatness. 
That is, we required $F$--flatness term by term 
in the superpotential,  
rather than allowing $F$--flatness to result from cancellation between terms. 
The absence of any non--zero terms from within $\vev{F_{\Phi_m}}$ and 
$\vev{W}$ is clearly sufficient to guarantee $F$--flatness along 
a given $D$--flat direction. 
However, such stringent demands are not necessary for $F$--flatness.
Total absence of all individual non--zero VEV 
terms can be relaxed: collections of such terms 
appear without breaking $F$--flatness, so long as the terms 
separately cancel among themselves in each $\vev{F_{\Phi_m}}$ and in $\vev{W}$. 
However, even when supersymmetry is retained at a given order in the superpotential 
via cancellation between several terms in a specific $F_{\Phi_m}$,
supersymmetry could well be broken at a slightly higher order.

Non--Abelian VEVs offer one solution to the stringent $F$--flatness issue.  
Because non--Abelian fields contain more than one field component, 
{\it self--cancellation} of a dangerous $F$--term can sometimes occur along 
\NA directions. That is, for some directions it may be possible to maintain
``stringent'' $F$--flatness even when dangerous 
$F$--breaking terms appear in the stringy superpotential. 
We will demonstrate self--cancellation of a \NA direction using, as examples,
the four \NA flat directions, FDNA1 through FDNA4, presented in Table I. 
These four directions are the simplest MSSM $D$--flat \NA directions that
are also $F$--flat to at least seventh order.
We will show that self--cancellation is not possible for FDNA1 and FDNA2, 
while it is for FDNA3 and FDNA4.
  
The singlet fields receiving VEVs, 
 $\{\p{12},\, \p{23},\, \pb{56},\, \p{4},\, \pp{4},\, \pb{4},\, \ppb{4},\,
   \Hs{31},\, \Hs{38}\},$
are the same for these four
flat directions.\footnote{For a list of the massless fields in
the FNY model see \cite{fny,cfn2}.}
The distinguishing aspect of these four directions is their \NA components. 
The \NA set for each direction is formed from a subset of 
the $SU(2)_H$ doublet fields $\{\H{23},\, \H{26},\, \V{40}\}$ 
and/or   
the $SU(2)'_H$ doublet fields $\{\H{25},\, \H{28},\, \V{37}\}$.
FDNA1 involves solely $SU(2)_H$ fields: $\H{23}$, $\H{26}$, and $\V{40}$, while
FDNA2 is a $SU(2)'_H$ parallel involving the
corresponding $\H{25}$, $\H{28}$, and $\V{37}$.
In contrast, FDNA3 and FDNA4 contain both $SU(2)_H$ and $SU(2)'_H$ 
doublets: the sets $\{\H{23}, \V{40}, \H{28}, \V{37}\}$, and
$\{\H{26}, \V{40}, \H{25}, \V{37}\}$, respectively.

Our four \NA flat directions can be separated into two sets, 
$\{$FDNA1, FDNA2$\}$ and $\{$FDNA3, FDNA4$\}$  
due to a global $Z_2$ symmetry under which all
$SU(2)_H$ and $SU(2)'_H$ fields are exchanged:
$\H{23}\leftrightarrow \H{25}$,
$\H{26}\leftrightarrow \H{28}$,
$\V{ 5}\leftrightarrow \H{ 9}$,
$\V{ 7}\leftrightarrow \H{10}$,
$\V{15}\leftrightarrow \H{19}$,
$\V{17}\leftrightarrow \H{20}$,
$\V{25}\leftrightarrow \H{29}$,
$\V{27}\leftrightarrow \H{30}$,
$\V{39}\leftrightarrow \H{35}$, and
$\V{40}\leftrightarrow \H{37}$.
This symmetry is maintained in the superpotential to very high (and probably all) order
in the superpotential. This implies that our findings regarding FDNA1 (FNDA3)
have parallels for FDNA2 (FDNA4).
Therefore we will examine only FDNA1 and FDNA3, but our  
findings will similarly apply to FDNA2 and FDNA4 after the appropriate field
exchanges.

Now let us focus on FDNA1. 
In this $D$--flat direction the ratio of norms of the VEVs is:  
\beqn
&&   \vert\vev{\p{12}}\vert^2=
   2 \vert\vev{\p{23}}\vert^2=
     \vert\vev{\pb{56}}\vert^2=
     \vert\vev{\Hs{31}}\vert^2=
     \vert\vev{\Hs{38}}\vert^2\\
&&  \phantom{\vert\vev{\p{12}}\vert^2}=
   2 \vert\vev{\H{23}}\vert^2=
   2 \vert\vev{\H{26}}\vert^2=
     \vert\vev{\V{40}}\vert^2
\equiv 2 \vert \vev{\alpha} \vert^2\,\, ;\quad {\rm and}
\label{fdsolval1b}\\
&& (\vert\vev{\p{4}}\vert^2+\vert\vev{\pp{4}}\vert^2)-
  (\vert\vev{\pb{4}}\vert^2+\vert\vev{\ppb{4}}\vert^2)=
\vert \vev{\alpha} \vert^2\,\, ,
\label{fdsolval1c}
\eeqn
where $\vev{\alpha}$ is the overall VEV scale determined
by eq.\ (\ref{daf}),
\beqn
\vev{\alpha}= \sqrt{\frac{g^2_s \MP^2 1344/112}{192 \pi^2}}
            \approx 1\times 10^{17}\,\, {\rm GeV}.
\label{alpdef}
\eeqn
This VEV ratio is fixed simply by the 
the Abelian $D$--terms (\ref{dtab},\ref{dtaban}) and the 
Cartan subalgebra (i.e., the diagonal) part of the $SU(2)_H$ and $SU(2)'_H$ 
$D$--terms.

\def\phm{\phantom{-}}
In generic \NA flat directions, 
the signs of the VEV components of a \NA field
are fixed by non--diagonal mixing of the 
VEVs in the corresponding \NA $D$-terms (\ref{dtgen}). 
Since FDNA1 contains $SU(2)_H$ doublets, we must require 
\beqn
\vev{D^{SU(2)_H}}=  
\vev{\H{23}^{\dagger} T^{SU(2)} \H{23} 
   + \H{26}^{\dagger} T^{SU(2)} \H{26}  
   + \V{40}^{\dagger} T^{SU(2)} \V{40}}= 0\,\, ,
\label{dfsu2}
\eeqn
where
\beqn
T^{SU(2)}\equiv \sum^{3}_{a= 1} T^{SU(2)}_a = 
\left ( 
\begin{array}{c c }
1 & 1-i \\
1+i & -1 
\end{array} \right )\,\, .
\label{TSU2}
\eeqn
The only solutions to (\ref{dfsu2}) 
consistent with
$\vert\vev{\H{23}}\vert^2= \vert\vev{\H{26}}\vert^2= \vert \vev{\alpha} \vert^2$
are
(up to a $\alpha \leftrightarrow -\alpha$ transformation)
\beqn
&&\vev{\H{23}} =\left ( 
\begin{array}{c}
\phm\alpha \\
-   \alpha
\end{array} \right )\, , \quad 
\vev{\H{26}} =\left ( 
\begin{array}{c}
\phm\alpha \\
-   \alpha
\end{array} \right )\,\, \quad 
\vev{\V{40}} =\left ( 
\begin{array}{c}
\phm\sqrt{2} \alpha \\
\phm\sqrt{2} \alpha
\end{array} \right )\, , 
\label{fdna1a} 
\eeqn
and
\beqn
&&\vev{\H{23}} =\left ( 
\begin{array}{c}
\phm\alpha \\
\phm\alpha
\end{array} \right )\, , \quad 
\vev{\H{26}} =\left ( 
\begin{array}{c}
\phm\alpha \\
\phm\alpha
\end{array} \right )\,\, \quad 
\vev{\V{40}} =\left ( 
\begin{array}{c}
\phm\sqrt{2} \alpha \\
-\sqrt{2}\alpha
\end{array} \right )\, . 
\label{fdna1b}
\eeqn

A ninth--order superpotential term,
$\p{23}\pb{56}\pp{4}\Hs{31}\Hs{38}\H{23}\H{26}\V{40}\V{39}$,
jeopardizes the 
$F$--flatness of this \NA $D$--flat direction via,  
\beqn
\vev{F_{\V{39}}} & \equiv& \vev{\frac{\partial W}{\partial \V{39}}}
\label{v35a}\\
                 & \propto & \vev{\p{23}\pb{56}\pp{4}\Hs{31}\Hs{38}}
                             \vev{\H{23}\cdot\H{26} \V{40} + 
                                  \H{23}\H{26}\cdot \V{40} +
                                  \H{26}\H{23}\cdot \V{40} }\, .\label{v35b}
\eeqn
Self--cancellation of this $F$--term could occur if the \NA VEVs resulted in 
\beqn
\vev{\H{23}\cdot\H{26} \V{40} +  \H{23}\H{26}\cdot \V{40} + \H{26}\H{23}\cdot \V{40} }=0\, .
\label{scna1}
\eeqn
However, neither (\ref{fdna1a}) nor (\ref{fdna1b}) produce this zero value.
Instead, they generate
\beqn
\vev{F_{\V{39}}} & = & -\lambda_9 \frac{ 8 {\alpha}^8}{\MP^6}
\left ( 
\begin{array}{c}
\pm 1 \\
+ 1
\end{array} \right )\, ,\label{v35d}
\eeqn
with $+1$ for (\ref{fdna1a}) and  $-1$ for (\ref{fdna1b}).  

In contrast to FDNA1, we now show that self--cancellation of a 
dangerous $F$--term 
{\it can} occur for FDNA3. That is, \NA $F$--term self--cancellation is 
consistent with $D$--term flatness for FDNA3.
Along the FDNA3 direction, the ratio of the singlet VEVs is the same as for FDNA1
except for the $\p{4}$ contribution. For FDNA3, 
\beqn
&& (\vert\vev{\p{4}} \vert^2+\vert\vev{ \pp{4}}\vert^2)-
   (\vert\vev{\pb{4}}\vert^2+\vert\vev{\ppb{4}}\vert^2)= 0 .
\label{fdsol3a}
\eeqn
The significant difference between FDNA3 and FDNA1 lies in FDNA3's \NA VEV ratio,
\beqn
\vert\vev{\H{23}}\vert^2  = 
\vert\vev{\V{40}}\vert^2  = 
\vert\vev{\H{28}}\vert^2  =  
\vert\vev{\V{37}}\vert^2  =  \vert \vev{\alpha} \vert^2\, .
\label{fdsolval3b}
\eeqn
The $SU(2)_H$ $D$--term,  
\beqn
\vev{D^{SU(2)_H}}=  
\vev{\H{23}^{\dagger} T^{SU(2)} \H{23} 
   + \V{40}^{\dagger} T^{SU(2)} \V{40}}= 0
\label{dfsu2sol}
\eeqn
has the two solutions  
\beqn
&&\vev{\H{23}} =\left ( 
\begin{array}{c}
\phm\alpha \\
-   \alpha
\end{array} \right )\, , \quad 
\vev{\V{40}} =\left ( 
\begin{array}{c}
\phm\alpha \\
\phm\alpha
\end{array} \right ), 
\label{fdna2a}
\eeqn
\no and
\beqn
&&\vev{\H{23}} =\left ( 
\begin{array}{c}
\phm\alpha \\
\phm\alpha
\end{array} \right )\, , \quad  
\vev{\V{40}} =\left ( 
\begin{array}{c}
\phm\alpha \\
-\alpha
\end{array} \right )\, .
\label{fdna2b}
\eeqn
The $SU(2)'_H$ $D$--term solutions for $\H{28}$ and $\V{37}$
have parallel form.

FDNA3's $F$--flatness is threatened by an eighth--order superpotential
term,
\beqn
\p{23}\pb{56}\Hs{31}\Hs{38}\H{23}\V{40}\H{28}\V{35}\, ,
\label{fdna3eo}
\eeqn
through 
\beqn
\vev{F_{\V{35}}} & \equiv& \vev{\frac{\partial W}{\partial \V{35}}}
\label{v35e}\\
                 & \propto & \vev{\p{23}\pb{56}\Hs{31}\Hs{38}}
                             \vev{\H{23}\cdot\V{40}}\vev{\H{28}}\label{v35f}.
\eeqn
Either set of $SU(2)_H$ VEVs (\ref{fdna2a}) or (\ref{fdna2b})
results in $\vev{\H{23}\cdot\V{40}}=0$. 
Hence $\vev{F_{\V{35}}} = 0$ self--cancellation is consistent with $D$--flatness
for FDNA3. Elimination of $\vev{F_{\V{35}}}$ makes FDNA3 flat to all finite order
in the superpotential.

One can show that generic self--cancellation of a dangerous $F$--term  
occurs when, for at least one \NA gauge group under which
some of the flat direction VEVs carry charge, 
the ratio of the powers of the corresponding fields in the $F$--term   
is equivalent to the ratio of the norms of the flat direction VEVs
carrying the given \NA charge.

\section{Hidden Sector Condensates}

Non--Abelian VEV directions such as FDNA3 or FDNA4 can yield a three generation
MSSM model while maintaining supersymmetry at the string /FI scale.
Supersymmetry must ultimately be broken slightly above the
electroweak scale, though. Along either of these two directions, our FNY model
shows qualitatively how supersymmetry may be broken dynamically by 
hidden sector field condensation after either of these two directions is invoked.
Recall that each of the flat directions FDNA3 and FDNA4 break 
both of the hidden sector $SU(2)_H$ and $SU(2)^{'}_H$ gauge symmetries, 
but leave untouched the hidden sector $SU(3)_H$. 
Thus, condensates of $SU(3)_H$ fields can initiate supersymmetry breaking
\cite{ln95}.

The set of nontrivial $SU(3)_H$ fields is composed of five triplets, 
\beqn
\{\H{42},\, \V{4},\, \V{14},\, \V{24},\, \V{34}\}\, ,
\label{strp}
\eeqn
and five corresponding anti--triplets,
\beqn
\{\H{35},\, \V{3},\, \V{13},\, \V{23},\,  \V{24}\}\, .
\label{satrp}
\eeqn
In both FDNA3 and FDNA4, singlet VEVs give unsuppressed FI--scale
mass to two triplets, $\V{24}$ and $\V{34}$,  
 and two anti--triplets, $\V{23}$ and $\V{33}$, 
via trilinear superpotential terms,\footnote{Note that these $SU(3)_H$ 
triplet mass terms also occur along the simplest MSSM singlet flat 
direction possible \cite{cfn1,cfn2}.}
\beqn
\vev{\p{12}} \V{23} \V{24} + \vev{\p{23}} \V{33} \V{34}
\label{trilt}
\eeqn
a slightly suppressed mass to one triplet/anti-triplet pair, $\H{42}$, 
and $\H{35}$, 
via a fifth order term, 
\beqn
\vev{\pb{56}\Hs{31}\Hs{38}} \H{42} \H{35}\,  .
\label{trilt2}
\eeqn
and a significantly suppressed mass to the pair,
$V{4}$ and $V{3}$,
via a tenth order term, 
\beqn
\vev{\p{23}\pb{56}\Hs{31}\Hs{38}\H{23}\V{40}\H{28}\V{37}} \V{4} \V{3}\,  .
\label{trilt3}
\eeqn
Before supersymmetry breaking, the last triplet/anti--triplet pair, 
$\V{14}$ and $\V{13}$, remain massless to all finite order. 

Consider a generic $SU(N_c)$ gauge group containing $N_f$ flavors 
of matter states in vector--like pairings 
$T_i \bar{T}_i$, $i= 1,\, \dots\, N_f$.
When $N_f < N_c$, the gauge coupling $g_s$, 
though weak at the string scale $\MS$, becomes strong
at a condensation scale defined by 
\beqn
\Lambda = \MP {\rm e}^{8 \pi^2/\beta g_s^2}\, ,
\label{consca}
\eeqn
where the $\beta$--function is given by,
\beqn
\beta = - 3 N_c + N_f\, .
\label{befn}
\eeqn
The $N_f$ flavors counted are only those that ultimately receive 
masses $m\ll \Lambda$.
Thus, for our model $N_c= 3$ and $N_f= 1$ 
(counting only the vector--pair, $\V{14}$ and $\V{13}$), which corresponds to  
$\beta = -8$ and results in an $SU(3)_H$ 
condensation scale
\beqn
\Lambda = {\rm e}^{-19.7}\MP \sim 7\times 10^9 \,\, {\rm GeV}. 
\label{consca2}
\eeqn

At this condensation scale $\Lambda$, the matter degrees of freedom are best
described in terms of the composite ``meson'' fields, $T_i \bar{T}_i$.
(Here the meson field is $\V{14}\V{13}$.)
Minimizing the scalar potential of our meson field induces
a VEV of magnitude,
\beqn
\vev{\V{14}\V{13}} = 
\Lambda^3 \left(\frac{m}{\Lambda}\right)^{N_f/N_c}\frac{1}{m}\, .
\label{ttv}
\eeqn
This results in an expectation value of
\beqn
\vev{W} = 
N_c \Lambda^3 \left(\frac{m}{\Lambda}\right)^{N_f/N_c}\, 
\label{wv}
\eeqn
for the non--perturbative superpotential.


Supergravity models are defined in terms of two functions,
the K\" ahler function, $G= K + {\rm ln}\, |W|^2$, where $K$ is the K\" ahler 
potential and $W$ the superpotential, and the gauge kinetic function $f$. 
These functions determine the supergravity interactions and the 
soft--supersymmetry
breaking parameters that arise after spontaneous breaking of supergravity, which is
parameterized by the gravitino mass $m_{3/2}$. The gravitino 
mass appears as a function of $K$ and $W$,
\beqn
m_{3/2} = \vev{{\rm e}^{K/2} W} .
\label{mkw}
\eeqn
Thus,
\beqn
m_{3/2} \sim \vev{{\rm e}^{K/2}} \vev{W} 
        \sim \vev{{\rm e}^{K/2}} N_c \Lambda^3 \left(\frac{m}{\Lambda}\right)^{N_f/N_c}\, . 
\label{mgeqa}
\eeqn
Restoring proper mass units explicitly gives,
\beqn
m_{3/2} \sim \vev{{\rm e}^{K/2}} N_c (\frac{\Lambda}{M_P})^3 
\left(\frac{m}{\Lambda}\right)^{N_f/N_c} M_P\, . 
\label{mgeqb}
\eeqn
Our meson field $\V{14}\V{13}$ 
will acquire a mass of at least the supersymmetry breaking
scale, so let us assume $m_{\V{14}\V{13}}\approx 1 $ TeV.
The resulting gravitino mass is 
\beqn
m_{3/2} &\sim& \vev{{\rm e}^{K/2}} 
\left(\frac{7\times 10^{9}\, {\rm GeV}}{2.4\times 10^{18}{\rm GeV}}
\right)^3 
\left(\frac{1000\, {\rm GeV}}{7\times 10^{9}\, {\rm GeV}}\right)^{1/3}
{2.4\times 10^{18}{\rm GeV}}
\nolabel\\  
&\approx& \vev{{\rm e}^{K/2}}\,  0.3\,\, {\rm eV} \, .
\label{mgeqd}
\eeqn

In standard supergravity scenarios, one generally obtains 
soft--supergravity--breaking parameters, such as scalar and gaugino masses and 
scalar interaction, that are comparable to the gravitino mass:
$m_o$, $m_{1/2}$, $A_o \sim m_{3/2}$.  
A gravitino mass of the order of the supersymmetry breaking scale 
would require $\vev{{\rm e}^{K/2}} \sim 10^{12}$ or $\vev{K}\sim 55$.
On the other hand, for a viable model, 
$\vev{{\rm e}^{K/2}}\sim {\cal{O}}(1)$ would 
necessitate a decoupling of local supersymmetry breaking (parametrized by
$m_{3/2}$) from global supersymmetry breaking (parametrized by
$m_{o}$, $m_{1/2}$). This is indeed possible in the context of no--scale
supergravity \cite{nl84}, endemic to weakly coupled string models.

In specific types of 
no--scale supergravity, the scalar mass $m_o$ and the scalar coupling
$A_o$ have null values thanks to the associated form of the K\" ahler 
potential. Furthermore, the gaugino mass can go as a power of 
the gravitino mass,
$m_{1/2} \sim \left(\frac{m_{3/2}}{\MP}\right)^{1-\frac{2}{3}q} \MP$,
for the standard no--scale form of $G$ and a non--minimal gauge kinetic
function $f\sim {\rm e}^{-A z^q}$, where $z$ is a hidden sector moduli field 
\cite{een84}.  
A gravitino mass in the range $10^{-5}$ eV $\lsim m_{3/2} \lsim 10^3$ eV
is consistent with the phenomenological requirement of $m_{1/2}\sim 100$ GeV 
for $\frac{3}{4}\gsim q \gsim \frac{1}{2}$.
Note that decoupling between the local and global breaking of supersymmetry
also appears to be realized in strongly coupled heterotic strings \cite{hora}.

\section{Discussion}

In this letter we have presented the four simplest \NA $D$--flat 
directions of the FNY model that (i) produce exactly the
MSSM fields as the only MSSM--charged fields in the low energy 
effective field theory and (ii) are flat to at least seventh order. 
$F$--flatness
for the first two directions is necessarily broken by
ninth order superpotential terms. 
For the last two directions, eighth order 
terms also pose a threat to $F$--flatness. 
All of these eighth and ninth order terms contain \NA
fields. 
For each of the latter two directions,
we showed that a set of \NA VEVs exist that is consistent 
with $D$--flat constraints and by which ``self--cancellation'' 
of the respective eighth order term can occur. 
By this, we mean
that for each specific set of \NA VEVs imposed by 
$D$--flatness constraints, 
the expectation value of the dangerous $F$--term is zero. 
Hence, the ``dangerous'' superpotential terms pose no problem
and our latter two directions become flat to all finite order. 
 
In \cite{cfn3} we discussed reasons why \NA VEVs are likely required 
for a phenomenologically viable
low energy effective MSSM, at least for the FNY string model.
Evidence has also been presented in the past suggesting this might be true 
as well for all MSHSM $Z_2\times Z_2$ models. This implies that there is
significant worth in exploring 
the generic properties of \NA flat directions in $Z_2\times Z_2$ models  
that contain exactly the MSSM three generations and two Higgs doublets
as the only MSSM--charged fields in the low energy effective field theory. 
For the next step in our study, we will
present in \cite{cfn5} a large set of systematically generated  
MSSM--producing \NA flat directions for the FNY model.
We will then analyze the phenomenological differences between 
our \NA directions and our past singlet directions.

\section{Acknowledgments}
This work is supported in part
by DOE Grants No. DE--FG--0294ER40823 (AF)
and DE--FG--0395ER40917 (GC,DVN,JW).
\newpage
\appendix

\section{Example Non-Abelian $D$-- and $F$--flat MSSM Directions} 

\def\ify{$\infty$ }
\def\ifw{${\infty}^{\ast}$ }

\def\x{ $\phantom{1}$}
\def\y{$\ast$}
\def\my{$\bar{\ast}$}
\def\ny{${(-)}\atop{\ast}$}

\begin{flushleft}
\begin{tabular}{|l||r||rrrrrr|rrr|rrr|}
\hline 
\hline
FD$\#$& Q'&\op{12}&\op{23}&\opb{56}&(\op{4})&\oHs{31}&\oHs{38}& \oH{23}&\oH{26}&\oV{40}&\oH{25}&\oH{28}&\oV{37} \\
\hline
FDNA1 &-1 &    2  &    1  &     2  &    1   &    2   &   2    &    1   &  1    &  2    &       &       &        \\
FDNA2 &-1 &    2  &    1  &     2  &    1   &    2   &   2    &        &       &       &   1   &   1   &   2    \\   
FDNA3 &-1 &    2  &    1  &     2  &    0   &    2   &   2    &    1   &       &  1    &       &   1   &   1    \\      
FDNA4 &-1 &    2  &    1  &     2  &    0   &    2   &   2    &        &  1    &  1    &   1   &       &   1    \\ 
\hline
\hline
\end{tabular}
\end{flushleft}
\no Table I: Example FNY directions flat through at least seventh order that 
contain VEVs of Non-Abelian charged Hidden Sector Fields.  
All component VEVs in these directions are uncharged under the MSSM gauge group.
Column one entries specify the class to which an example direction belongs.
Column two entries give the anomalous charges $Q'\equiv Q^{(A)}/112$ of the 
flat directions.
The next several column entries specify the ratios of the norms of the VEVs. 
The $\p{4}$--related component is the net value (in units of the square 
overall VEV scale) of
$\mvev{\Phi_4} + \mvev{\Phi^{'}_4} - \mvev{\bar{\Phi}_4} - \mvev{\bar{\Phi}^{'}_4}$. 
E.g., a ``1'' in the $\Phi_4$ column for FDNA1 specifies that   
$\mvev{\Phi_4} + \mvev{\Phi^{'}_4} - \mvev{\bar{\Phi}_4} - \mvev{\bar{\Phi}^{'}_4} = 
1\times \mvev{\alpha}$. 

\begin{flushleft}
\begin{tabular}{|l||l|l|}
\hline 
\hline
FD$\#$& Dangerous $W$ Terms & Self--Cancellation $F$--Flatness Solution ?\\
\hline
FDNA1 & $ \p{23} \pb{56} \pb{4} \Hs{31} \Hs{38} \H{23} \H{26} \V{40} \V{39}$ & No. \\
FDNA2 & $ \p{23} \pb{56} \pp{4} \Hs{31} \Hs{38} \H{25} \H{28} \V{37} \V{35}$ & No. \\
FDNA3 & $ \p{23} \pb{56}        \Hs{31} \Hs{38} \H{23} \V{40} \H{28} \V{35}$ & Yes, via
          $\{\vev{\H{23}}, \vev{\V{40}}\}\, .$ \\ 
FDNA4 & $ \p{23} \pb{56}        \Hs{31} \Hs{38} \H{26} \V{40} \H{25} \V{35}$ & Yes, via
          $\{\vev{\H{26}}, \vev{\V{40}}\}\, .$ \\
\hline
\hline
\end{tabular}
\end{flushleft}

\no Table II: Dangerous $F$-breaking superpotential terms for flat directions in Table I.
 
\no Column one entries specify the class of a flat direction.
The entry in the next column specifies superpotential terms that can (possibly) break $F$--flatness
and the last column entry indicates whether or not there is an allowed set of \NA VEVs, consistent with $D$-flatness constraints,
 by which $F$--flatness may be maintained through self--cancellation. 
\hfill\vfill\eject

               
\vfill\eject

\bigskip
\medskip

\def\bibiteml#1#2{ }
\bibliographystyle{unsrt}

\hfill\vfill\eject
\end{document}